\documentclass[showpacs,reprint,twocolumn,preprintnumbers,amsmath,amssymb,aps,prl,superscriptaddress]{revtex4-1}

\usepackage{graphicx}
\usepackage{dcolumn}
\usepackage{bm}
\usepackage[normalem]{ulem}
\usepackage{subfigure}
\usepackage{color}
\usepackage{times}

\renewcommand{\eqref}[1]{(\ref{#1})}

\begin{document}
\title{Optimal and Robust Quantum Metrology Using Interaction-Based Readouts} 
\author{Samuel~P.~Nolan}
\email{uqsnolan@uq.edu.au}
\affiliation{School of Mathematics and Physics, The University of Queensland, Brisbane, Queensland, Australia}
\author{Stuart~S.~Szigeti}
\affiliation{School of Mathematics and Physics, The University of Queensland, Brisbane, Queensland, Australia}
\affiliation{ARC Centre of Excellence for Engineered Quantum Systems, University of Queensland, Brisbane, Queensland, Australia}
\affiliation{Department of Physics, Centre for Quantum Science, and Dodd-Walls Centre for Photonic and Quantum Technologies, University of Otago, Dunedin 9010, New Zealand}
\author{Simon~A.~Haine}
\affiliation{Department of Physics and Astronomy, University of Sussex, Brighton, United Kingdom}

\date{\today}

\begin{abstract}
Useful quantum metrology requires nonclassical states with a high particle number and (close to) the optimal exploitation of the state's quantum correlations. Unfortunately, the single-particle detection resolution demanded by conventional protocols, such as spin squeezing via one-axis twisting, places severe limits on the particle number. Additionally, the challenge of finding optimal measurements (that saturate the quantum Cram{\'e}r-Rao bound) for an arbitrary nonclassical state limits most metrological protocols to only moderate levels of quantum enhancement. ``Interaction-based readout'' protocols have been shown to allow optimal interferometry \emph{or} to provide robustness against detection noise at the expense of optimality. In this Letter, we prove that one has great flexibility in constructing an optimal protocol, thereby allowing it to also be robust to detection noise. This requires the full probability distribution of outcomes in an optimal measurement basis, which is typically easily accessible and can be determined from specific criteria we provide. Additionally, we quantify the robustness of several classes of interaction-based readouts under realistic experimental constraints. We determine that optimal \emph{and} robust quantum metrology is achievable in current spin-squeezing experiments.

\end{abstract}


\maketitle
 
Nonclassical states enable precision measurements below the shot-noise limit (SNL) \cite{Caves1981, Wineland:1992}. However, despite many proof-of-principle experiments \cite{Schnabel:2010, Giovannetti:2011, LIGO:2013, Taylor:2016, Pezze2016}, 
a useful (i.e., high-precision) quantum-enhanced measurement has yet to be performed. This is partially due to the fragility of nonclassical states to typical noise sources \cite{DemkowiczDobrzanski2015} 
and the difficulty in marrying quantum-state-generation protocols with the practical requirements of high-precision metrology \cite{Robins:2013, Matthews:2016}; addressing these issues is an active research area \cite{Haine2013, Szigeti:2014b, Dur:2014, Tonekaboni2015, Haine2015, Haine2015a, Unden:2016, Kruse:2016}. A key limitation is detection noise \cite{Pezze2007, Kardynal2008, Bakr:2009, Zhang2012, Hume2013, Bohnet2014, Pezze2016, Sun2016}, which makes $n$ and $n \pm \sigma$ particles indistinguishable. Quantum-enhanced measurements typically require single-particle resolution ($\sigma \sim 1$), which restricts them to small particle numbers, since the requisite counting efficiency rapidly becomes unattainable as particle number increases.

\begin{figure*}
\includegraphics[width=\textwidth]{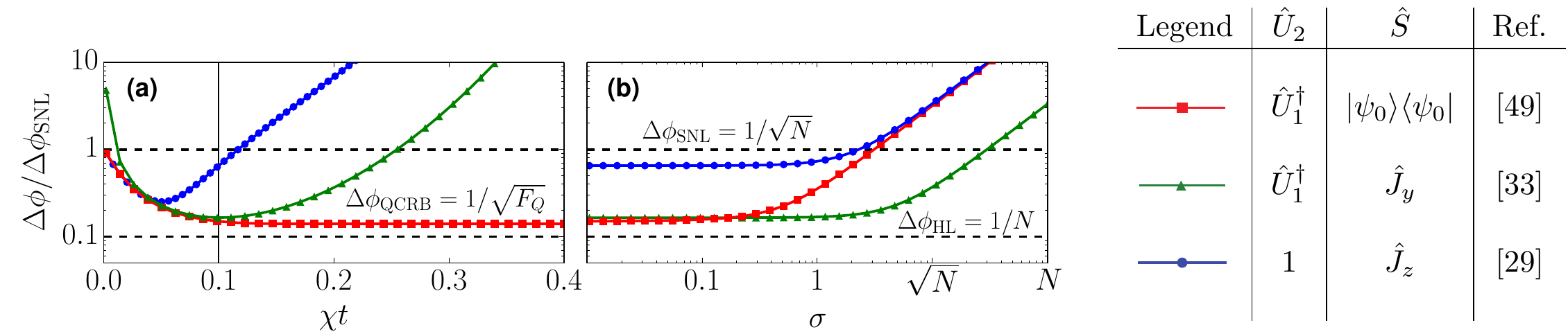} 
\caption{Phase sensitivity (normalized to the SNL) of state $|\psi_\phi\rangle$ [Eq.~(\ref{eq:echo})] for conventional spin squeezing (i.e. the trivial protocol $\hat{U}_2=1$, blue circles), compared to an optimal protocol (red squares) and a robust protocol (green triangles), all with $\hat{U}_1$ an OAT interaction. \textbf{(a)} Sensitivity at different squeezing strengths $\chi t$ for perfect particle detection ($\sigma=0$). The trivial protocol quickly reaches the ENGS (`over-squeezed') regime, and although the robust protocol also provides enhanced sensitivity it does not saturate the QCRB. The QCRB, identical for all schemes, is saturated with an echo followed by a measurement that projects onto the initial state (i.e. that counts instances of maximal $J_x$, red squares) or, as shown below, for an arbitrary parity-conserving readout with a full spin-resolving measurement. \textbf{(b)} Dependence of sensitivity on detection noise $\sigma$ for fixed $\chi t=0.1$, indicated by vertical line in \textbf{(a)}. An echo followed by an average pseudospin measurement (green triangles) is significantly more robust than the other schemes, which both require detection resolution at the single-particle level.}
\label{fig:gain}
\end{figure*}

Another challenge is that many protocols are suboptimal, as they do not fully exploit the state's quantum correlations. 
Specifically, an estimate of classical parameter $\phi$ obtained from measurement signal $\hat{S}$ has a precision $\Delta \phi^2= \min_\phi \text{Var}[\hat{S}(\phi)]/[\partial_\phi \langle \hat{S}(\phi) \rangle ]^2$. A quantum-enhanced estimate surpasses the SNL $\Delta \phi^2 = 1 / N$ for particle number $N$, however it is only optimal if it saturates the quantum Cram{\'e}r-Rao bound (QCRB) $\Delta \phi^2 = 1/F_Q$, where $F_Q$ is the quantum Fisher information (QFI) \cite{Caves1994, Paris2009, Toth2014, DemkowiczDobrzanski2015}. For example, consider the nonclassical $N$-qubit states generated via the one-axis twisting (OAT) Hamiltonian \cite{Kitagawa1993, Gross2010, Riedel:2010, Leroux:2010}. 
Typical spin-squeezing procedures use the expectation of pseudospin as the signal, yielding a minimuim sensitivity $\Delta \phi^2 \sim N^{-5/3}$. However, OAT can produce entangled non-Gaussian states (ENGS), which can achieve the Heisenberg limit (HL) $F_Q=N^2$ and therefore have enormous metrological potential. Nevertheless, for ENGS an average pseudospin estimator yields precision \emph{worse} than the SNL [Fig.~\ref{fig:gain}(a)].

One pathway to either optimal (saturates the QCRB) or robust (against detection noise) quantum metrology is so-called ``interaction-based readouts'' which take the form
\begin{equation} \label{eq:echo}
	|\psi_\phi \rangle= \hat{U}_2 \hat{U}_\phi \hat{U}_1 |\psi_0 \rangle,
\end{equation}
where $|\psi_0\rangle$ is the initial (unentangled) state, $\hat{U}_1$ the entangling operation (e.g., OAT), $\hat{U}_\phi$ the phase encoding, and $\hat{U}_2$ the interaction-based readout applied prior to measurement. These protocols can provide significant robustness to detection noise and give improved sensitivity \cite{Davis2016, Frowis2016, Hosten2016, Anderson:2016, Anderson:2017, Davis:2017} - although a protocol that is both optimal and robust has remained illusive. 
Specifically, echo protocols \cite{Yurke1986, Leonhardt:1994, Toscano2006, Goldstein2011, Jing:2011, Marino:2012, Hudelist:2014, Ma:2015, Gabbrielli2015, Chen2015, Davis2016, Macri2016, Davis:2017, Linnemann2016, Manceau:2017, Szigeti:2017} which perfectly time reverse the first entangling unitary ($\hat{U}_2 = \hat{U}_1^\dag$) and then project onto the initial state have been shown to saturate the QCRB for arbitrary pure states $\hat{U}_1|\psi_0 \rangle$ \cite{Macri2016} (red squares Fig.~\ref{fig:gain}). However, this scheme is not robust to detection noise. In contrast, an echo followed by a measurement of the average pseudospin provides robustness, but does not saturate the QCRB \cite{Davis2016, Davis:2017} (green triangles Fig.~\ref{fig:gain}). 

In this Letter, we demonstrate that both optimal \emph{and} robust protocols are possible. Using the classical Fisher information (CFI) we show that accessing the full probability distribution of measurement outcomes in a particular (usually easily accessible) basis saturates the QCRB. Crucially, these measurements remain optimal for the large class of readouts $\hat{U}_2$ that conserve parity with respect to this basis, which means one is free to choose a $\hat{U}_2$ suitable for \emph{any} other purpose, including improved robustness to detection noise.
We investigate several readouts and confirm that echoes provide significant robustness, although readouts that lack time-reversal symmetry can be similarly or more robust. For situations where the state preparation time is a fixed resource, we show that echoes are \emph{never} optimal for short OAT times - which is the operating regime for current experiments. 

\textit{Criteria for optimal interferometry.---}Suppose $\phi$ is encoded onto state $\hat{\rho}$ via unitary $\hat{U}_\phi = \exp(-i \phi \hat{G})$. Subsequent measurements in some orthonormal basis $\{|m \rangle \}$ allow $\phi$ to be estimated from the probabilities $P_m(\phi)=\langle m| \hat{U}_{\phi} \hat{\rho} \hat{U}_\phi^\dagger|m \rangle$. The estimate's precision is bounded by the Cram{\'e}r-Rao bound $\Delta \phi^2 \geq 1/F_C(\phi)$ where $F_C$ is the CFI, which relates to the probabilities via the Hellinger distance
\begin{equation} \label{eq:dH}
d_H^2(\phi_1,\phi_2) \equiv 1-\textstyle\sum_m \sqrt{P_m(\phi_1) P_m(\phi_2)},
\end{equation}
since $d_H^2(0,\phi) = F_C(0) \phi^2/8 + \mathcal{O}(\phi^3)$  \cite{Strobel2014, Pezze2016, CFI_defn}.

In general, there is no guarantee that the CFI associated with this measurement is optimal. However, we prove the CFI \emph{always} saturates the QCRB if: 
\begin{enumerate}
	\item the input state is a parity eigenstate \cite{Campos2003}: $\hat{\Pi} \hat{\rho} = (-1)^p \hat{\rho}$ with $p = 0, 1$ for $\hat{\Pi} = \sum_m (-1)^m |m \rangle \langle m |$;
	\item the generator $\hat{G}$ \emph{flips} parity (i.e., $\hat{\Pi} \hat{G} \hat{\Pi} = -\hat{G}$). 
\end{enumerate}
In principle, this holds for most spin-squeezing interferometry experiments and SU(1,1) interferometers \cite{Yurke1986}.

We proceed by expanding $P_m(\phi)=P_m(0)+\phi P'_m(0) + \phi^2 P''_m(0)/2+\mathcal{O}(\phi^3)$, where $P'(0)=i ( \langle m | \hat{\rho} \hat{G} |m \rangle - c.c )$ and $P''(0)=\langle m| \hat{G} \hat{\rho} \hat{G} |m \rangle - \langle m| \hat{\rho} \hat{G}^2 |m \rangle + c.c$. Conditions (1) and (2) imply that $P'_m(0) = 0$, since $\langle m | \hat{G} \hat{\rho} |m \rangle = - \langle m | \hat{G} \hat{\rho} |m \rangle = 0$. Furthermore, $P_m(0) = (-1)^{m+p} \langle m | \hat{\rho} |m \rangle$ and $\langle m | \hat{G} \hat{\rho} \hat{G} |m \rangle = (-1)^{m+p+1} \langle m | \hat{G} \hat{\rho} \hat{G} |m \rangle$, hence $P_m(0) \langle m| \hat{G} \hat{\rho} \hat{G} |m \rangle = 0$.  After a binomial expansion of the square root in Eq.~(\ref{eq:dH}),
\begin{equation} \label{Hell_simp}
	d_H^2(0,\phi) - \mathcal{O}(\phi^3) = \phi^2 \tfrac{\langle \hat{G}^2 \rangle_{\hat{\rho}}}{2}  = \phi^2 \tfrac{F_C(0)}{8},
\end{equation}
where $\langle \hat{G}^2 \rangle_{\hat{\rho}} \equiv \text{Tr} \{ \hat{G}^2 \hat{\rho}\}$. Finally, our two assumptions ensure $\langle \hat{G}\rangle_{\hat{\rho}}=0$, implying $\text{Var}(\hat{G}) = \langle \hat{G}^2 \rangle_{\hat{\rho}}$. Equating powers of $\phi$ in Eq.~(\ref{Hell_simp}) gives $F_C(0)=4 \text{Var}(\hat{G})$. Since $F_C \leq F_Q \leq 4 \text{Var}(\hat{G})$ \cite{Toth2014}, then $F_C(0)=F_Q$, proving that our measurement is optimal if conditions (1) and (2) hold.

This is \emph{not} simply a proof that the QCRB is saturable. Rather, it concretely determines the optimal measurement basis \cite{paritybasis} (typically easily accessible), without the tedious or impossible requirement of diagonalizing the symmetric logarithmic derivative. Crucially, it also shows that including a second unitary $\hat{U}_2$ after the phase-encoding, such that $P_m(\phi)=\langle m| \hat{U}_2  \hat{U}_{\phi}\hat{\rho} \hat{U}_\phi^\dagger  \hat{U}_2 ^\dagger |m \rangle$, leaves the CFI unchanged provided $\hat{U}_2$ conserves parity with respect to the measurement basis. This means that, fundamentally, a readout protocol is unnecessary: all parity-conserving interaction-based readouts have identical CFI, and are equivalent to simply doing nothing after the phase encoding ($\hat{U}_2 = 1$). Indeed, all three schemes in Fig.~\ref{fig:gain}(a), which have wildly-different phase sensitivities and experimental complexities, can saturate the QCRB if a full probability distribution is used. Of course, robustness to detection noise still requires a non-trivial $\hat{U}_2$.

\textit{One-axis twisting interferometry.---}A broad class of interferometry is possible within two-bosonic-mode systems of $N$ particles. Provided $N$ is fixed, these systems can be described by the SU(2) algebra $[\hat{J}_i, \hat{J}_j] = i \epsilon_{ijk} \hat{J}_k$, where $\epsilon_{ijk}$ is the Levi-Civita symbol \cite{Yurke1986}. Spin-squeezing protocols, which quantum enhance the state prior to phase encoding, are described within this framework. The `trivial' protocol in Fig.~\ref{fig:gain} is: (1) spin squeezing generated via OAT, $\hat{U}_1 = \exp[-i \hat{J}_x \theta(N,\chi t)] \exp(-i \hat{J}_z^2 \chi t) \equiv \hat{U}_\text{OAT}(t)$, where $\theta(N,\chi t)$ is a rotation angle that minimizes $\text{Var}(\hat{J}_z)$ \cite{Kitagawa1993}; (2) phase-encoding via Mach-Zehnder interferometry $\hat{U}_\phi = \exp(-i \phi \hat{J}_y)$; (3) measurement of population difference $\hat{S}=\hat{J}_z$.
Other spin-squeezing protocols include two-axis twisting \cite{Kitagawa1993} and the ``twist-and-turn" scheme. \cite{Strobel2014, Muessel2015}. 

If the initial state is a maximal $\hat{J}_x$ eigenstate (a spin-coherent state), then its parity with respect to the $\hat{J}_x$ eigenbasis remains unchanged under any of these spin-squeezing protocols. Passing the resultant nonclassical state through a Mach-Zehnder ($\hat{G} = \hat{J}_y$) and making measurements in the $\hat{J}_x$ eigenbasis satisfies conditions~(1) and (2), implying via our above result that the CFI saturates the QCRB, thereby attaining the best phase sensitivity.

Spin squeezing has been demonstrated in trapped ions \cite{Meyer2001, Monz2011, Britton:2012}, Bose-Einstein condensates (BECs) \cite{Esteve2008, Riedel2010, Berrada2013}, cold atoms in cavities \cite{Leroux2010, SchleierSmith2010, Muessel2015, Schmied2016}, and optical systems \cite{Dong:2008, Corney:2008, Ono2016}, and has enhanced proof-of-principle interferometric measurements \cite{Gross2010, Ockeloen2013, Hosten2016}, including atomic clocks \cite{Appel:2009, Leroux2010b} and magnetometers \cite{Sewell2012, Muessel2014}. Note the `proof-of-principle' aspect to these experiments; spin squeezing has not yet resulted in a useful measurement that surpasses current shot-noise-limited high-precision devices. This is due to the fragility of spin-squeezed states, which has limited the degree of squeezing and/or particle numbers to modest values. Maximizing the metrological benefits of squeezing, preferably with minimal increases in experimental complexity, is clearly desirable. Our above result suggests that estimating the phase by constructing the full probability distribution, (rather than from an estimate of the mean value of the psuedospin \cite{Davis:2017} or the probability of a single outcome \cite{Macri2016}), could help achieve this goal. 


\begin{figure}
\includegraphics[width=\columnwidth]{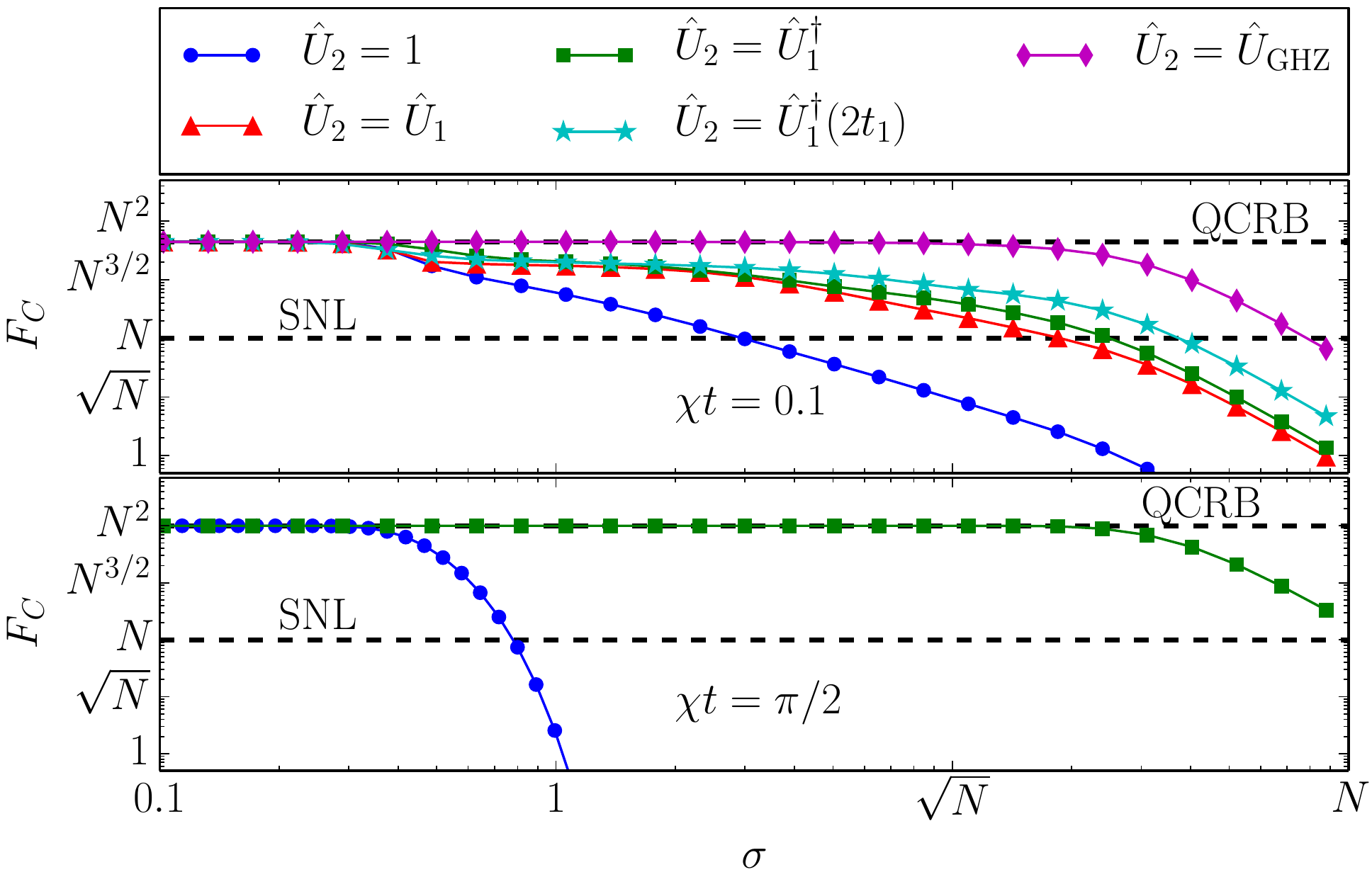} 
\caption{Variation of $\mathrm{max}_\phi F_C$ with detection noise for a state produced by $\hat{U}_1=\hat{U}_{\mathrm{OAT}}$ and readout $\hat{U}_2$ in the optimal basis with \textbf{(top)} $\chi t = 0.1$ and \textbf{(bottom)} a GHZ state with $\chi t =\pi/2$, where an echo gives the HL with detection noise exceeding $\sqrt{N}$. Here $N=100$.}
\label{fig:Fcvsloss}
\end{figure}

\textit{Robustifying against detection noise.---} After determining the optimal measurement basis with conditions (1) and (2), the full probability distribution in this basis must be estimated. A spin-resolving measurement can give this information, as reported in \cite{Strobel2014}. Although \emph{perfect} spin-resolving measurements render echoes unnecessary, detection noise makes this difficult to achieve in practice, and so interaction-based readouts will still play an important role in optimal parameter estimation. We investigate the CFI when detection noise is present, and although we confirm that echoes can provide significant robustness to detection noise, we show that better sensitivities are possible with non-echo protocols.

For concreteness, consider the nonclassical state generated by evolving a maximal $\hat{J}_x$ eigenstate under OAT for time $t$. After passing through a Mach-Zehnder, interaction-based readout $\hat{U}_2$ is applied (leaving the QFI unchanged) and a spin-resolving measurement made in the optimal basis $\{ | m\rangle \}$. We model detection noise in this measurement as discrete Gaussian noise $G_{m}(\sigma)$ of variance $\sigma^2$, corresponding to an uncertainty $\sigma$ in the measured particle number. This noise distorts the measured probabilities (and consequently the CFI), which we account for by replacing $P_m(\phi)$ with the \emph{conditional}
probabilities $\tilde{P}_m(\phi|\sigma)=\sum_{m'} C_{m'} G_{m-m'}(\sigma) P_{m'}(\phi)$ \cite{Smerzi2013, Gabbrielli2015} where $C_{m'}=\{\sum_m G_{m-m'} \}^{-1}$ normalizes $G_{m-m'}$. This is still a `spin-resolving' measurement, as it returns (imperfect) information about the full distribution $P_m$ (in contrast to an estimate of the distribution mean).

\begin{figure}
\includegraphics[width=\columnwidth]{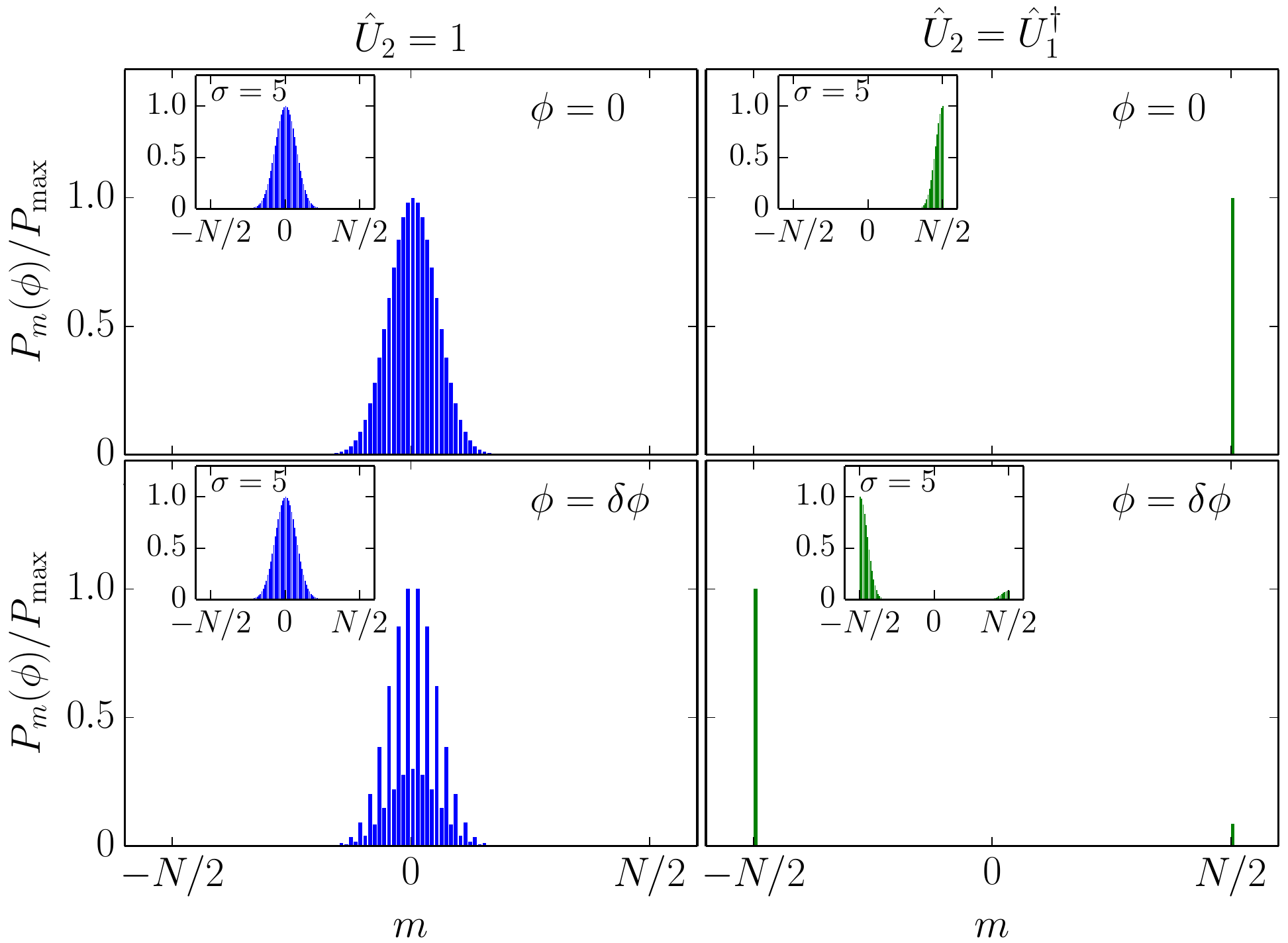} 
\caption{$P_m(\phi)$ histograms for a $N=100$ GHZ state in the optimal basis with an echo (\textbf{right}, green, optimal basis is $\hat{J}_y$), and without an echo (\textbf{left}, blue, optimal basis is $\hat{J}_x$). The \textbf{top} and \textbf{bottom} panels differ by a small rotation $\exp(-i \hat{J}_y \delta \phi)$ ($\delta \phi = N^{-1/2}$). Inset histograms of $\tilde{P}_m(\phi|\sigma)$ are for the same state with detection noise $\sigma^2 = N/4$. The greater the distinguishability of the distributions after rotation $\delta \phi$, the larger the CFI.}
\label{fig:noisycat}
\end{figure}

\begin{figure*}
\includegraphics[width=\textwidth]{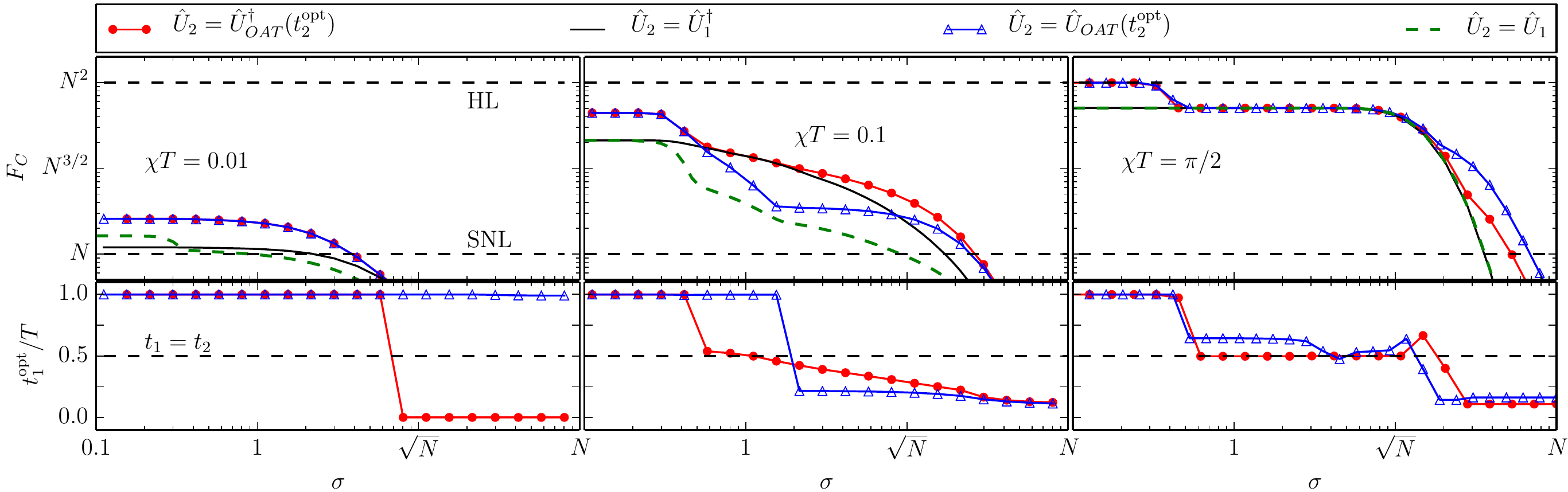} 
\caption{For $N=100$ particles with $\hat{U}_1=\hat{U}_{\mathrm{OAT}}(t_1)$ we fix total squeezing time $T=t_1+t_2$ for various $T$. We explore two time-asymmetric readouts: $\hat{U}_2=\hat{U}_{\mathrm{OAT}}^\dagger(t_2)$ (solid red circles) and ``pseudo-echo" $\hat{U}_2=\hat{U}_{\mathrm{OAT}}(t_2)$ (empty blue triangles). (\textbf{top}) Comparison of maximum $F_C$ (at optimum $\phi$) of each protocol to the special cases ($t_1=t_2$) of an echo (solid black) and ``pseudo-echo" (dashed green). 
(\textbf{bottom}) The optimal $t_1$ that maximizes the CFI \cite{local_opt}. 
}
\label{fig:fixedT}
\end{figure*}

In Fig.~\ref{fig:Fcvsloss} we plot the CFI for various interaction-based readouts $\hat{U}_2$. As expected, an echo ($\hat{U}_2 = \hat{U}_1^\dag$) provides significant robustification over no echo ($\hat{U}_2 = 1$). This robustness is \emph{not} achieved by the echo proposed in \cite{Macri2016} (red squares Fig.~\ref{fig:gain}), which only accesses the maximal $\hat{J}_x$ component ($m=N/2$) rather than the full probability distribution $P_m$.
However, we find a class of \emph{time-asymmetric} protocols capable of outperforming echoes. Specifically, if $\hat{U}_1 = \hat{U}_\text{OAT}(t_1)$ corresponds to OAT evolution of duration $t_1$, then $\hat{U}_2 = \hat{U}_\text{OAT}^\dagger(t_2)$ with $t_2>t_1$ generally outperforms an echo ($t_2=t_1$) provided squeezing strength $\chi t_1$ is modest. 

Robustness to detection noise can also be achieved with ``pseudo-echoes'', $\hat{U}_2=\hat{U}_1$, which do not reverse the time evolution of $\hat{U}_1$ \cite{Frowis2016, Hosten2016}. Although less effective than echoes or asymmetric time-reversal protocols, pseudo-echoes nevertheless provide good robustification, and are an excellent alternative when time reversal is difficult or impossible. For example, the interatomic collisions that generate many-body entanglement in BECs can only be reversed by changing the inter- and intra- component couplings \cite{Haine2014}. This typically requires a Feshbach resonance \cite{Inouye:1998} unavailable to many atomic species, and even if possible is limited to small condensates and squeezing durations due to inherent instabilities in attractive condensates \cite{Berge:2000, Wuster:2007} or instabilities and poor mode-matching in two-component mixtures \cite{Haine2014, Lee:2016}. Implementing echoes in soliton-based atom interferometers \cite{McDonald:2014, Helm:2015, Haine:2017} and optical fibers \cite{Dong:2008, Corney:2008, Andersen:2016} is similarly impractical.

For OAT, the creation of a GHZ state \cite{Bouwmeester:1999} provides an upper limit on the QFI (i.e. the HL $F_Q = N^2$), since at $\chi t_2>\pi/2$ the state revives towards the initial condition (a maximal $\hat{J}_y$ eigenstate). The most robust readout is $\hat{U}_2 = \hat{U}_\text{OAT}(\chi t_2=\pi/2) \equiv \hat{U}_{\mathrm{GHZ}}$, with a spin-resolving measurement in the $\hat{J}_y$ basis, since this projects onto the initial state. Although such a protocol is infeasible in current experiments, this extreme case provides insight into why these protocols successfully robustify OAT to detection noise. Figure~\ref{fig:noisycat} (left, blue histograms) shows a GHZ state with $\hat{U}_2=1$ in the optimal measurement basis (the $\hat{J}_x$ eigenbasis), before and after a small perturbation $\delta \phi$. The states at $\phi = 0$ and $\phi = \delta \phi$ are distinguished only by a decay of odd $P_m$, obscured by even a small amount of detection noise, making the two distributions virtually indistinguishable and resulting in a small CFI. In contrast, a GHZ state followed by an echo (i.e. at $\phi=0$ it has returned to the initial state, and so the optimal basis here is the $\hat{J}_y$ eigenbasis) retains a large Hellinger distance (large CFI) even in the presence of significant detection noise. Astonishingly, a GHZ state provides sensitivity at the HL for detection noise \emph{exceeding} $\sqrt{N}$ [Fig. \ref{fig:Fcvsloss} (bottom)].

\textit{Optimal protocols with total-time constraint.---}The overall duration of OAT experiments is limited by particle losses and/or dephasing \cite{Li2008, Li2009, Riedel:2010} and a desire to maintain high repetition rates. Therefore, in an experiment restricted to some fixed total squeezing time $T=t_1+t_2$ there is potentially a trade-off between increasing $t_1$ in order to increase the QFI via $\hat{U}_1(t_1)$ and increasing $t_2$ in order to optimally tune the readout $\hat{U}_2(t_2)$. This is explored in Fig.~\ref{fig:fixedT}, which plots the maximum $F_C$ (top) and corresponding $t_1$ (bottom) for small, medium, and large $T$.

For sufficiently small detection noise, any interaction-based readout (i.e., $t_2 > 0$) confers no benefit (consistent with our proof above), and the best strategy is to simply maximize the state's quantum correlations (and therefore QFI) by choosing $t_1=T$. In contrast, for large $T$ (e.g., a GHZ state with $\chi T=\pi/2$) and non-negligible detection noise an echo remains the best strategy up until $\sigma \sim \sqrt{N}$. The reason is simple: when evolving an initial maximal $\hat{J}_x$ eigenstate under OAT, the QFI quickly reaches a plateau at $N^2/2$ [Fig.~\ref{fig:gain}(a)]. Thus, an echo remains optimal, as there is no trade-off between increasing the QFI via $ t_1$ and increasing the robustness via $t_2$. In this large-$T$ regime pseudo-echoes perform as well as echoes. 

Figure~\ref{fig:fixedT} (middle) shows regimes where it is beneficial to choose time-asymmetric readouts such as $\hat{U}_2=\hat{U}_{\mathrm{OAT}}^\dagger(t_2)$ over echoes, although protocols without time-reversal [e.g. $\hat{U}_2=\hat{U}_{\mathrm{OAT}}(t_2)$] perform poorly. 

The experiment \cite{Gross2010} used $\chi T \approx 0.01$ (and $N=170$). For fixed, small squeezing times on this order [Fig.~\ref{fig:fixedT} (left)] the optimal strategy is $\hat{U}_2=1$ (no readout), even for modest detection noise. This is the operating regime for most current spin-squeezing experiments.

\textit{Conclusions.---} 
We have shown that constructing the full probability distribution in the optimal measurement basis [i.e. one that satisfies conditions (1) and (2)] yields a phase estimate that saturates the QCRB. Crucially, this is true for any parity-conserving readout, including one that provides robustness to detection noise (such as an echo), enabling both optimal \emph{and} robust quantum metrology. Consequently, nonclassical states such as ENGS, which are not traditionally useful for spin squeezing, could enhance future metrological devices, and the single-particle detection requirements that limit other protocols to small particle numbers (e.g. \cite{Macri2016}) could be relaxed.

We also showed that if the total spin-squeezing duration is fixed and short, an echo gives poorer results than simply squeezing for longer, even for considerable detection noise. Furthermore, we have found a class of asymmetric time-reversal protocols superior to echoes, and also shown that pseudo-echoes, which do not require any time reversal, provide comparable robustness. Pseudo-echoes are advantageous for interferometers that use BECs, bright-solitons, or optical fibers, where it is difficult or impossible to time-reverse the state's evolution.  These results give additional flexibility in protocol design, and could find near-term applications in current short-duration spin-squeezing experiments.

\emph{Acknowledgements.---}We thank Joel Corney and Jacob Dunningham for invaluable discussions and assistance. Numerical simulations were performed on the University of Queensland School of Mathematics and Physics computing cluster ``Dogmatix,'' with thanks to I.~Mortimer for computing support. S.P.N. acknowledges support provided by an Australian Postgraduate Award. S.S.S. acknowledges the support of the Australian Research Council Centre of Excellence for Engineered Quantum Systems (Project No.~CE110001013), the Australia Awards-Endeavour Research Fellowship, and the Dodd-Walls Centre for Photonic and Quantum Technologies. S.A.H. has received funding from the European Union's Horizon 2020 research and innovation programme under the Marie Sklodowska-Curie Grant Agreement No.~704672.

\bibliographystyle{apsrev4-1}

\bibliography{echo_bib_v3}

\begin{thebibliography}{88}
\bibitem [{loc()}]{local_opt}%
  \BibitemOpen
  \href@noop {} {}\bibinfo {howpublished} {$t_1^\text{opt}$ are not guaranteed
  global optima, since there exist many locally-optimal $t_1$ with similar
  $F_C$.}\BibitemShut {Stop}%
\end{thebibliography}

\end{document}